\documentclass[12pt]{iopart}
\usepackage{graphicx}
\usepackage{subfigure}
\usepackage{amsmath}

\begin{document}
\title{Weak boson production measured in PbPb and pp collisions by CMS}

\author{Jorge A. Robles (for the CMS collaboration)}

\address{University of California, Davis. Physics Department, Davis CA, 95616 }

\ead{robles@physics.ucdavis.edu}

\begin{abstract}
The unprecedented center-of-mass energy available at the LHC offers unique opportunities for studying the properties of the strongly-interacting QCD matter created in PbPb collisions at extreme temperatures and very low parton momentum fractions. %With its high precision, large acceptance for tracking and calorimetry, and a trigger scheme that allows the analysis of almost each minimum bias PbPb event by the high-level trigger, CMS is fully equipped to measure muons and electrons in the high multiplicity environment of nucleus-nucleus collisions.
Electroweak boson production is an important benchmark process at hadron colliders. Precise measurements of Z production in heavy-ion collisions can help to constrain nuclear PDFs as well as serve as a standard candle of the initial state in PbPb collisions at the LHC energies. The inclusive and differential measurements of the Z boson yield in the muon decay channel will be presented, establishing that no modification is observed with respect to next-to-leading order pQCD calculations, scaled by the number of incoherent nucleon-nucleon collisions. The status of the Z measurement in the electron decay channel, as well as the first observation of W$\rightarrow\mu\nu$ in heavy ion collisions will be given. The heavy-ion results will be presented in the context of those obtained in pp collisions with the CMS detector.%\cite{CMSdet}. 
\end{abstract}

\section{Introduction}

With the increase of center-of-mass energy at the LHC, electroweak probes are accessible for the first time in heavy-ion collisions. Electroweak bosons are of interest because they go unaffected  through the hot and dense matter often refered to as the quark gluon plasma (QGP). The study of the W and Z in their lepton and dilepton channels, respectively, will yield a clean probe of the initial state for high-$q^{2}$ processes. The Z boson decaying in the dilepton channel can therefore be used as a standard candle. The precise measurement of this process can help constrain cold nuclear matter (CNM) effects, such as shadowing. %With these measurements one can study CNM effects in PbPb without the need for pA collisions for the Z case. 
The high precision, large acceptance for tracking and calorimetry, and high efficiency trigger scheme allows CMS to measure muons and electrons in the high multiplicity environment of nucleus-nucleus collisions\cite{CMSdet}.

\section{W $\rightarrow\mu  \nu$ in PbPb collisions }

Inclusive W production can be measured via a high-transverse-momentum ($p_{T}$) muon, combined with a missing-transverse-energy (MET) signature in the opposite side in azimuth. Muons that pass a loose isolation cut and have a $p_{T}^{\mu}>$ 20 GeV/c are selected. Figure 1 (Left) shows the $p_{T}^{\mu}$ distribution of single muons. A visible enhancement in the range 25-50 GeV/c can be seen, compatible with a leptonic W-decay signature. The neutrino's $p_{T}$ is inferred from the MET of the event. The vectorial sum of all the tracks, with $p_{T} > $2 GeV/c, should be balanced with the  the missing transverse energy of the event, which should be the transverse energy of the neutrino. The W transverse mass, $M_{T}$, is shown in Fig 1 (Right) for those events with an opening angle cut $|\Delta\phi^{\mu\nu}| >$ 2 radians. The data points are shown in full circles, and are overlaid with a realistic W Monte Carlo (MC) simulation embedded in HYDJET\cite{Hydjet} PbPb events. A reasonable agreement is found between data and simulation after all the analysis cuts have been applied.

\begin{figure}[ht]
\centering
\subfigure{
\includegraphics[width=13pc]{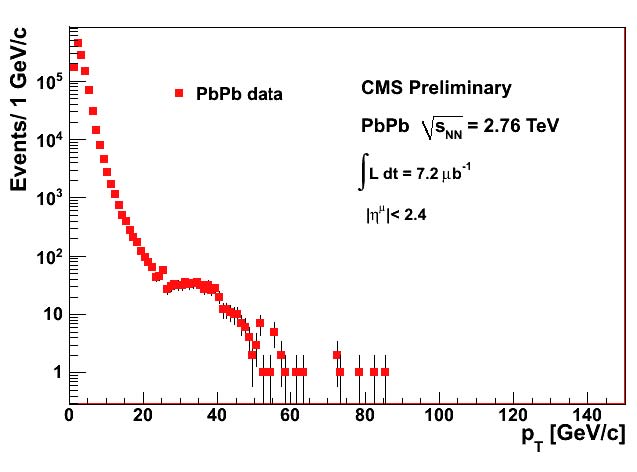}
\label{fig:WmuonPt}
}
\hspace{3pc}
\subfigure{
\includegraphics[width=14pc]{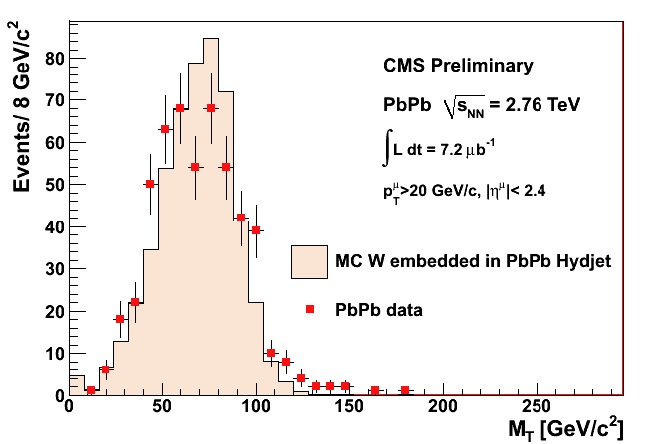}
\label{fig:WTmass}
}
\label{Wimages}
\centering
\caption{(Left) Single muon $p_{T}$, CMS PbPb data.  (Right) Transverse mass of W candidates, CMS PbPb data with MC W events. The simulation has been scaled to match the data. }
\end{figure}

\section{Z $\rightarrow e^{+} e{^-}$ in PbPb collisons}

 Z production is an important benchmark process in pp and PbPb collisions. The measurement in the dilepton channel offers many advantages, given that the decay leptons do not interact via the strong force and can traverse the QGP unaffected. However, there are some experimental challenges that arise from the reconstruction of the $e^{+} e^{-}$ pairs in PbPb collisions.  The total energy collected, due to the underlying event,  in the electromagnetic calorimeter($|\eta| < 3.0$) changes as a function of centrality, and affects the electron energy used to reconstruct the Z invariant mass. Figure 2 (Left) shows a simulation of the reconstructed invariant mass  as a function of the event centrality. The empty circles are obtained with the standard clustering used in pp collisions; a strong dependence on centrality is observed. The empty squares are obtained using an optimized clustering algorithm for PbPb collisions; where a decrease in centrality dependency can be seen. The mass dependence with centrality can be used to calibrate the detector response for electron and photon analyses using the Z mass as a benchmark. Figure 2 (Right) shows the Z invariant mass in the $e^{+} e^{-}$ channel obtained from the PbPb data. The Z candidates require each electron of an $e^{+} e^{-}$ pair to have  $p_{T}^{e} >$10 GeV/c and $|\eta^{e}| < 2.4$. We integrated a yield of 27 candidates.

\begin{figure}[ht]
\centering
\subfigure{
\includegraphics[width=13pc]{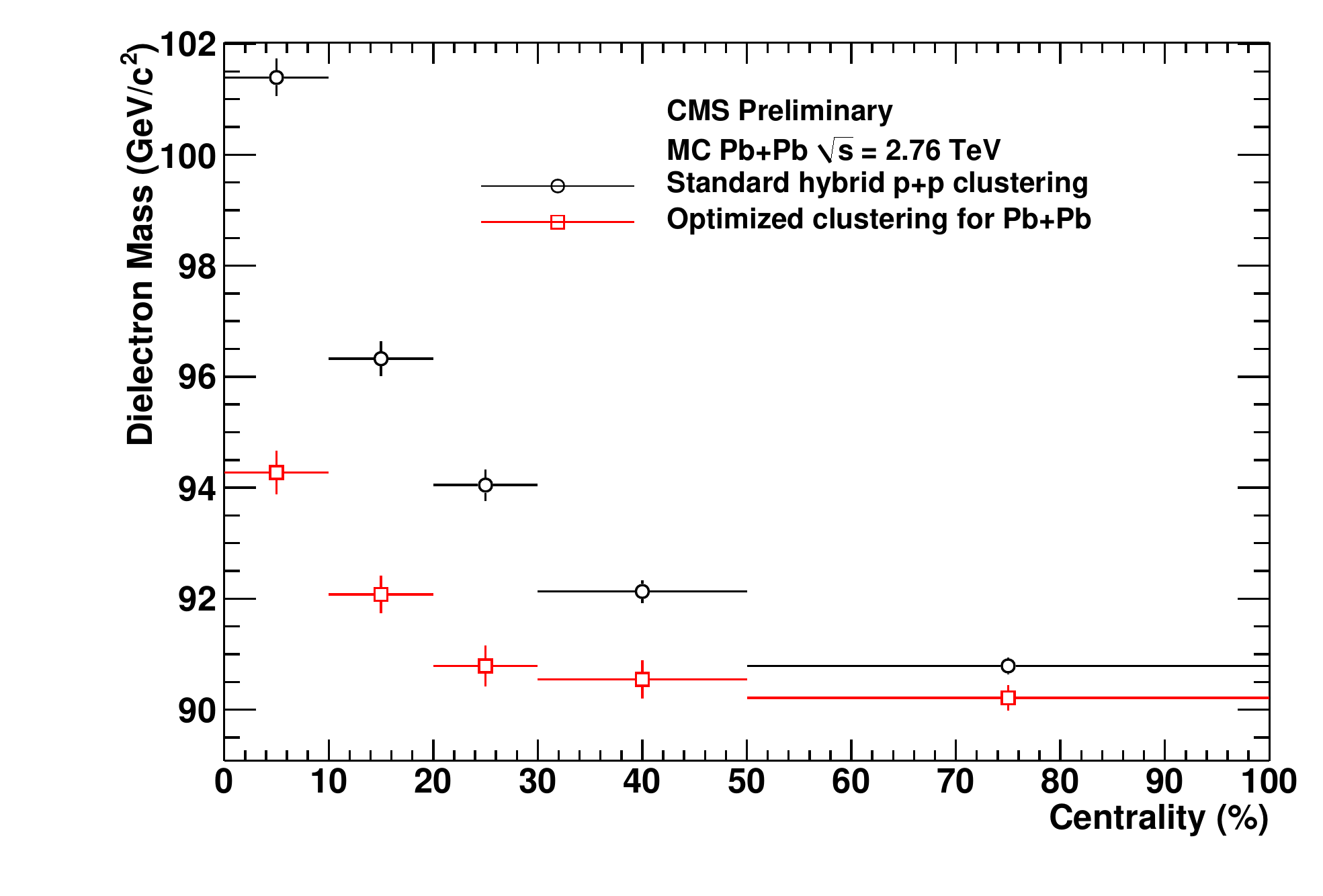}
\label{fig:ZMassCent}
}
\hspace{3pc}
\subfigure{
\includegraphics[width=14pc]{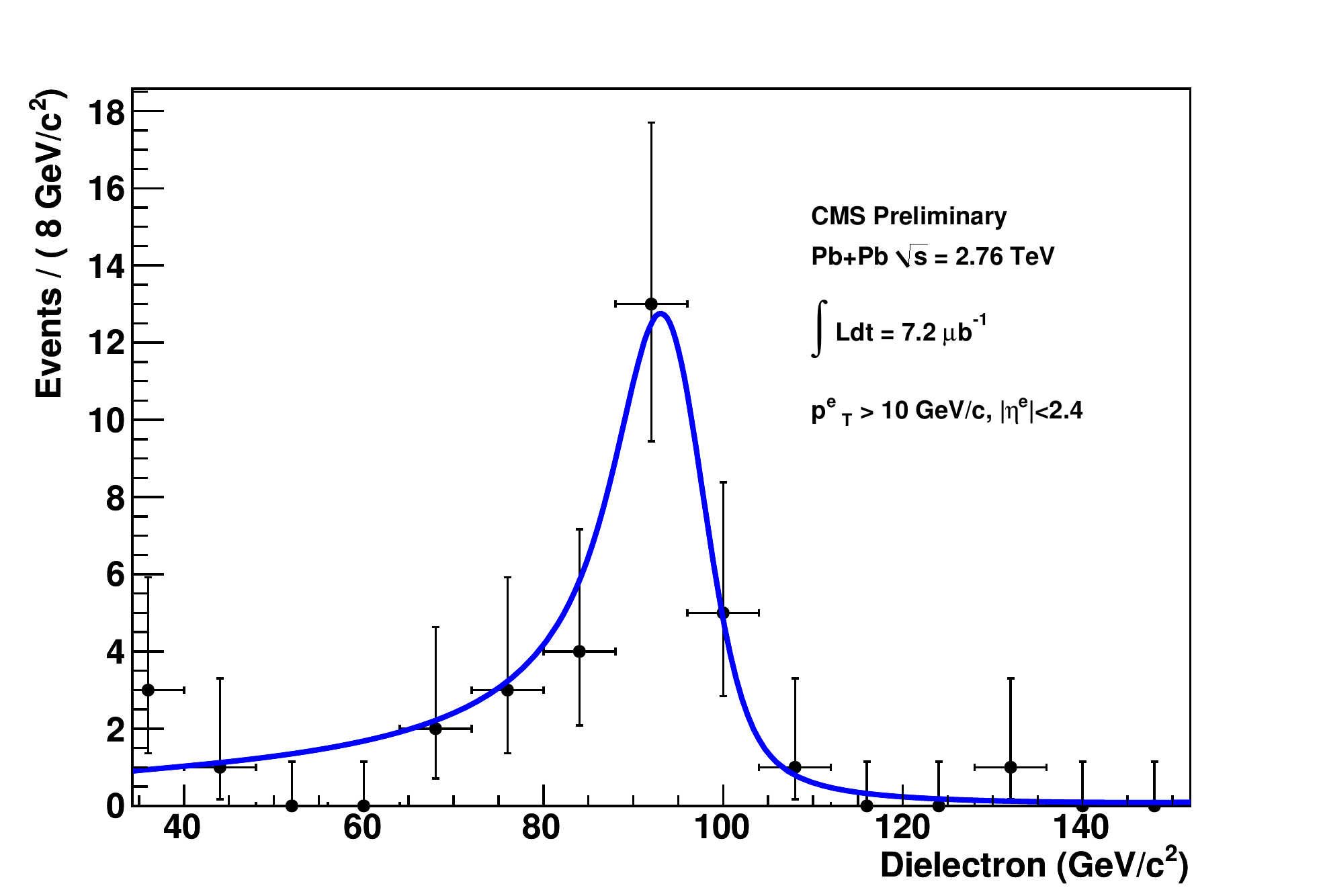}
\label{fig:ZmassEE}
}
\label{ZeePlots}
\caption{(Left) Reconstructed Z mass in $e^{+} e^{-}$ channel as a function of centrality.   (Right) Z invariant mass in $e^{+} e^{-}$ channel, fitted with a crystal-ball function.}
\end{figure}

\section{Z $\rightarrow \mu^{+} \mu{^-}$ Results in PbPb collsions}

Among the leptonic decays of the electroweak bosons, the study of the Z  in the  $\mu^{+} \mu{^-}$ channel is the cleanest one experimentally. The Z is produced from hard collisions via $q\bar{q} \rightarrow Z$ process. Once produced, it decays within the medium on a time scale of $\approx 10^{-24}$ seconds. The dimuon decay channel is of special interest because the Z, as well as the muons, traverse the medium unaffected by the strongly interacting QGP. Given that the entire process is virtually untouched by the medium, it offers a probe to measure initial state effects. 

Predictions\cite{Kart,Ramona,Zhang,Paukkunen,Neufeld} indicate that the effects that can modify the Z boson production in heavy-ion collisions are rather small. Shadowing is expected to have the largest contribution $\approx$10 - 20$\%$ \cite{Paukkunen}. The isospin effect arises from the difference in quark content of protons and neutrons that collide in the PbPb case as opposed to pp, and it is expected to contribute less than 3$\%$ \cite{Paukkunen}. Finally, scattering and energy loss of the initial partons should have an effect on the order of 2$\%$\cite{Neufeld}.

\begin{figure}[ht]
\centering
\includegraphics[width=14pc]{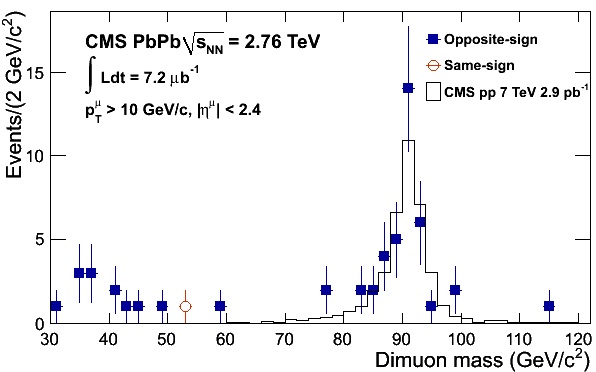}
\caption{\label{ZMassPbPb} Z invariant mass distribution in PbPb collisions at $\sqrt{s}$ = 2.76 TeV in the dimuon channel.}
\end{figure}

The Z $\rightarrow \mu^{+} \mu{^-}$ heavy-ion results by CMS are detailed in \cite{ZPbPb}. The reconstruction of the Z boson in the dimuon channel is done by requiring two opposite-charge muons, each with $p_{T}^{\mu} > $ 10 GeV/c and $|\eta^{\mu}| < 2.4$. A loose set of quality cuts applied to each of the muons is enough to extract a clear peak in the 60 - 120 GeV/$c^{2}$ mass region. Figure 3 shows the invariant mass spectrum from opposite-charge muon pairs in full squares. The only same-sign pair is also shown. The data points are overlaid with a histogram from pp collision data normalized to an integral of 39 counts.  The structure found in the 30 - 50 GeV/$c^{2}$ region is due to the dimuon continuum from other physics processes (mainly $b\bar{b}$ production).  A visible agreement between the PbPb data points and the pp histogram data is indicative of a comparable detector performance in pp and heavy-ion collisions.

Figure 4 shows the differential yields as a function of $p_{T}^{Z}$(Left), $y^{Z}$(Center) and $N_{part}$(Right). In all cases the PbPb data agrees, within uncertainties, with a pp POWHEG simulation scaled by the nuclear geometry. Within measurement uncertainties none of the subtler effects can be discerned. The main feature is a flat distribution as a function of centrality. This is expected for a probe that is unmodified by the presence of the dense QGP produced in central collisions.

\begin{figure}[ht]
\centering
\subfigure{
\includegraphics[width=11pc]{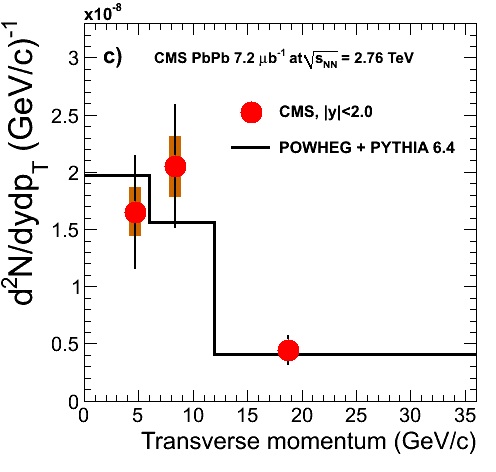}
\label{fig:ZPT}
}
\hspace{0pc}
\subfigure{
\includegraphics[width=11pc]{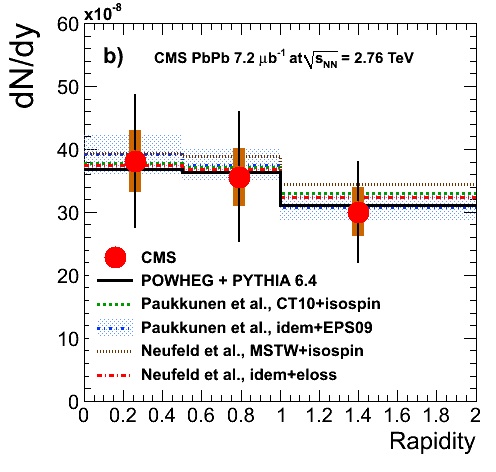}
\label{fig:ZrapPbPb}
}
\hspace{0pc}
\subfigure{
\includegraphics[width=11pc]{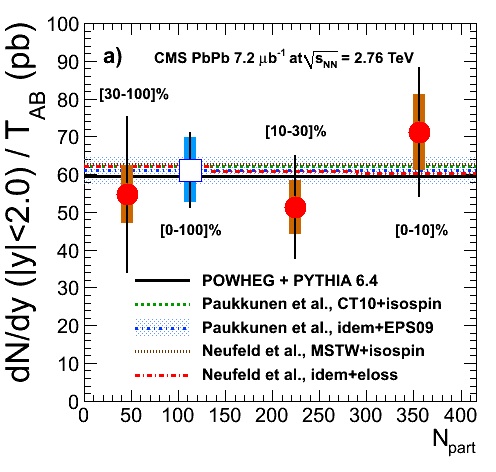}
\label{fig:ZRaaPbPb}
}
\label{ZmumuPlots}
\caption{(Left) Z differential yield as a function of $p_{T}$.  (Center) Z differential yield as a function of rapidity. (Right) Z differential yield divided by the nuclear overlap function as a function of $N_{part}$ obtained from Glauber model.}
\end{figure}

\section{Conclusion}
A first observation of W$\rightarrow\mu\nu$ and Z $\rightarrow e^{+} e{^-}$ in heavy-ion collisions by CMS is presented. The Z $\rightarrow \mu^{+} \mu{^-}$ yield is measured as function of centrality, $p_{T}^{Z}$  and $y^{Z}$. It is found to scale with the nuclear geometry as compared to a pp simulation.  An $R_{AA}$ of 1.00 $\pm$0.16(stat) $\pm$0.14(syst) is obtained assuming a d$\sigma_{pp}$/dy = 59.6 pb obatined with POWHEG. This confirms the expectation that the Z is unmodified by the hot QGP.

\end{document}